\documentstyle[titlepage,12pt]{article}
\newfont{\bg}{cmr10 scaled\magstep3}
\newcommand{\gsimeq}
{\hbox{ \raise3pt\hbox to 0pt{$>$}\raise-3pt\hbox{$\sim$} }}
\newcommand{\lsimeq}
{\hbox{ \raise3pt\hbox to 0pt{$<$}\raise-3pt\hbox{$\sim$} }}

\newcommand{\plb}[3]{Phys. Lett. {\bf B#1} (#2) #3} 
\newcommand{\prl}[3]{Phys. Rev. Lett. {\bf #1} (#2) #3}
\newcommand{\prd}[3]{Phys. Rev. {\bf D#1} (#2) #3}
\newcommand{\npb}[3]{Nucl. Phys. {\bf B#1} (#2) #3}
\newcommand{\npbps}[3]{Nucl. Phys. {\bf B}(Proc. Suppl.) {\bf #1} (#2) #3}
\newcommand{\prog}[3]{Prog. Theor. Phys. {\bf #1} (#2) #3}

\newcommand{\be}{\begin{eqnarray}}
\newcommand{\ee}{\end{eqnarray}}

\newcommand{\delpq}{\delta_{pq}}
\newcommand{\gfive}{\gamma_5}

\newcommand{\tr}{\mbox{\rm tr}}

\begin{document}
\begin{titlepage}
\rightline{}


\addtocounter{footnote}{1}

\begin{center}

{\large 
\hfill
\parbox{5cm}{\normalsize KUNS-1517 HE(TH)~98/11\\UT-822 
\\{\tt  hep-lat/9806013}}\\
\vspace{1.5cm}
\bf 
Weak coupling expansion of massless QCD  with 
\\
\vspace{0.5cm}
overlap Dirac operator and axial $U(1)$ anomaly 
}

\vspace{0.5cm}

\vspace{2cm}
{\sc Yoshio Kikukawa}\footnote{e-mail address:
kikukawa@gauge.scphys.kyoto-u.ac.jp}
\\
\vspace{.5cm}
{\it Department of Physics, Kyoto University, Kyoto, 606-8502, Japan}
\vspace{1cm}

{\sc Atsushi Yamada}\footnote{e-mail address:
atsushi@hep-th.phys.s.u-tokyo.ac.jp} 
\\
\vspace{.5cm}
{\it Department of Physics, University of Tokyo, Tokyo, 113 Japan}
\vspace{2cm}

\date{\normalsize June, 1998}

{\bf ABSTRACT}
\end{center}
We discuss the weak coupling expansion of massless QCD 
with the Dirac operator which is derived by Neuberger based on
the overlap formalism and satisfies the Ginsparg-Wilson relation. 
The axial $U(1)$ anomaly associated to the chiral transformation 
proposed by L\"uscher is calculated as an application and is shown 
to have the correct form of the topological charge density
for perturbative backgrounds. The coefficient of the anomaly 
is evaluated as a winding number related to a certain 
five-dimensional fermion propagator.
\end{titlepage}

\baselineskip = 0.7cm


Recently Dirac operators which may describe exactly massless
fermion on a lattice are proposed 
by Neuberger \cite{neu1} 
based on the overlap formalism \cite{overlap,odd-dim-overlap}
and by Hasenfratz, Laliena and Niedermayer \cite{ha} based on the
renormalization group method 
\cite{fixed-point-action-YM,fixed-point-action-QCD}.
These Dirac operators are very different from each other but both 
satisfy the Ginsparg-Wilson relation \cite{gw}, which ensures that 
the fermion propagator anti-commutes with $\gfive$ at non-zero 
distances and thus respects chiral symmetry in that sense. 
Subsequently L\"uscher observed \cite{lu} that the Ginsparg-Wilson 
relation implies an exact symmetry of the fermion action and the 
anomalous behavior of the fermion partition function under its 
flavor-singlet transformation is expressed in terms of the index 
of the Dirac operator arised as the Jacobian factor of the path
integral measure, providing a clear understanding of 
the exact index theorem on a lattice in Ref. \cite{ha}. 

Though it was pointed out in Ref. \cite{gw} that Dirac operators 
satisfying the Ginsparg-Wilson relation will necessarily exhibit non-local 
behavior in the presence of the dynamical gauge fields, a recent
perturbative analysis of chiral gauge theory in the overlap 
formalism \cite{ay} suggests some of them will be well controllable 
as a field theory on a lattice. 
 
In this paper we consider the weak coupling expansion of massless QCD 
with the Dirac operator which is derived by Neuberger and 
satisfies the Ginsparg-Wilson relation. Then we apply our results 
to analyze the axial $U(1)$ anomaly following the formulation 
of Ref. \cite{lu} and confirm that the Jacobian factor is given in the 
form of the topological charge density for slowly varing perturbative 
gauge fields, 
supplementing the original calculation of the same quantity of Ref. \cite{gw}. 

The covariant anomaly in the vacuum overlap formalism has been discussed 
by Neuberger and Narayanan \cite{neuberger-geometric-anomaly}, 
Randjbar-Daemi and Strathdee \cite{daemi-strathdee} and Neuberger 
\cite{neuberger-geometric-anomaly}. Note also that the axial anomaly
has been calculated in the context of the domain-wall QCD by 
Shamir \cite{shamir-anomaly}.


We begin with a brief review of the symmetry argument of Ref. \cite{lu}. 
The partition function we consider is of the form 
\begin{eqnarray} 
Z=\int d\psi d \bar{\psi} e^{-a^4 \Sigma \bar{\psi} D\psi} 
\label{eqn:z}
\end{eqnarray}
with a lattice Dirac operator $D$ satisfying the Ginsparg-Wilson relation 
\begin{eqnarray}
D\gfive + \gfive D= a D \gfive D.  
\label{eqn:gw}
\end{eqnarray}
In Ref. \cite{lu} L\"uscher pointed out that with the aid of the 
relation (\ref{eqn:gw}) the fermion action 
$a^4 \Sigma \bar{\psi} D\psi$ is invariant under the infinitesimal 
transformation $\psi \rightarrow \psi + i \epsilon T \delta \psi $, 
$\bar{\psi} \rightarrow \bar{\psi} + i \epsilon \delta \bar{\psi}T $
where $T$ is a generator of the rotation in the flavor space and    
\begin{eqnarray}
\delta \psi= \gfive(1-\frac{1}{2} aD) \psi, \,\,\,\,\,
\delta \bar{\psi} =\bar{\psi}(1-\frac{1}{2} aD) \gfive .
\label{eqn:lusym}
\end{eqnarray}
The path integral measure 
yields the Jacobian factor $-i\epsilon a \tr \{ T \gfive D \} $, which
does not vanish only for the flavor singlet chiral rotation, 
accounting for the index theorem of Ref. \cite{ha}. 

A Dirac operator satisfying the relation (\ref{eqn:gw}) was proposed 
by Neuberger in Ref. \cite{neu1} and now we review its brief derivation  
starting from the domain wall fermion of the Shamir type \cite{shamir} 
based on Ref. \cite{neu3}, trying to clarify the physics backgrounds of
the Dirac operator. 

The domain wall fermion of this type is a Wilson fermion in five dimensions 
with the finite size $N_5$ of the fifth space and is described 
by the action 
\begin{eqnarray} 
S &=& \sum_{s,t} a^4 \sum_{m,n} \bar{\psi}(m,s) D_5(\mu)_{ms,nt}\psi(n,t), 
\label{eqn:shamia} \\
D_5(\mu)_{ms,nt}  &=&        
\sum_{\mu=1}^{4} \gamma_\mu C_\mu (m,n) \delta_{s,t} 
+B_5(m,n) \delta_{s,t} 
-\frac{1}{a_5} P_L \delta_{s+1,t} 
-\frac{1}{a_5} P_R \delta_{s,t+1} 
\nonumber \\
& &+\mu  P_L  \delta_{s,N} \delta_{t,1} 
+\mu  P_R  \delta_{s,1} \delta_{t,N}, 
\\
C_\mu (m,n) &=& \frac{1}{2a} \Bigl[  
\delta_{m+\mu,n} U_\mu (m) 
-\delta_{m,n+\mu} U^\dagger_\mu (n) 
\Bigr],\,\,\,
B_5 (m,n) = \frac{1}{a_5} +B (m,n), 
\\
B (m,n)  &=& \frac{M_0}{a} +\frac{r}{2a} \sum_\mu
\Bigl[ 2\delta_{m,n}-  \delta_{m+\mu,n} U_\mu (m) 
- \delta_{m,n+\mu} U^\dagger_\mu (n) 
\Bigr],
\end{eqnarray}
where $m,n$ and $a$ denote the four dimensional space indices and their 
lattice spacing while $s,t$ and $a_5$ denote 
the indices and the lattice spacing of the fifth dimension 
($1 \leq s,t \leq N_5 $). 
In the action Eq.~(\ref{eqn:shamia}), the link variables 
$U_\mu (m)$ couple only between the fields $\bar{\psi}(m,s)  $ 
and $\psi(m \pm \mu,s)$ so that there is no gauge interaction in the 
fifth space. The terms proportional to $1/a_5$ come from the kinetic and 
Wilson terms in the fifth direction and the Wilson parameter in that 
direction is set to be unity. 
We have omitted the factor $a_5$ in front of the sum $\sum_{s,t} $ 
so that $\psi$ has the correct dimension $3/2$ for fermions in the 
four dimensions. The role of the parameter $\mu$ is explained later. 

The action Eq.~(\ref{eqn:shamia}) is also regarded as the action of the $N_5$ 
flavor Wilson fermions in four dimensions with a specific mass matrix.  
With a suitable choice of $M_0$ the action describes one Dirac fermion 
with the mass $\mu+ {\cal O}( e^{- N_5} /a) $ and $N_5-1$ 
Dirac fermions with the mass of the order of the inverse lattice 
spacing \cite{ex1}. Then the following partition function 
\begin{eqnarray} 
& &\int d \bar{\psi} d\psi \int d\bar{\psi}_{PV}    d \psi_{PV} 
e^{-a^4  \Sigma \bar{\psi} D_5(0) \psi + 
a^4 \Sigma \bar{\psi}_{PV} D_5(1/a_5)\psi_{PV} } 
\nonumber \\
& &=\det D_5(\mu=0) /\det D_5(\mu=\frac{1}{a_5})   
\label{eqn:pv}
\end{eqnarray}    
may describe one massless Dirac fermion regulated {\`a la } Pauli-Villars 
by one Dirac fermion with the mass $1/a_5$ if the $N_5-1$ heavy 
fermions cancel out. Such cancellation allows to take the limit 
$N_5 \rightarrow \infty$. Then taking the limit $a_5 \rightarrow 0$, 
one massless Dirac fermion remains. 

Now we compute the determinant $\det D_5(\mu) $ and obtain the final 
expression of Eq. (\ref{eqn:pv}) as a single determinant, which allows  
the path integral expression like eq. (\ref{eqn:z}). 
In the chiral basis of the $\gamma$-matrices defined by
\begin{equation}
 \gamma_\mu= \left( \begin{array}{cc} 0 & \sigma_\mu \\
                                       \sigma_\mu^\dagger & 0 
                      \end{array} \right) , \quad 
 \gamma_5 = \left( \begin{array}{cc} -1 & 0 \\
                                      0 & 1
                      \end{array} \right) , \qquad 
\sigma_\mu= ( -i \sigma_1,-i \sigma_2, -i \sigma_3, 1 ) ,
\end{equation}
the Dirac field is written in the chiral components as 
\begin{equation}
\psi = \left( \begin{array}{c} \psi_L \\ \psi_R \end{array} \right) , 
\quad \bar \psi = \left( \bar \psi_R , \bar \psi_L \right) .
\end{equation}
We also introduce the notation 
\begin{equation}
 \psi^\dagger = \bar \psi \gamma_4 ,
\end{equation}
althoug it does not mean hermitian conjugate.
Then the action can be written as  
\begin{eqnarray}
S&=&\sum_{s,t} a^4 \sum_{m,n} 
( {\psi}^\dagger_L(m,s),\,\,  {\psi}^\dagger_R(m,s) )    
\Bigl[ 
\left(
\begin{array}{cc}       
C^\dagger & B_5 \\ 
B_5       & -C  \\
\end{array}
\right)
\delta_{st}
+ 
\left(
\begin{array}{cc}       
0      & 0 \\ 
-1/a_5 & 0 \\
\end{array}
\right)
\delta_{t,s+1}
\nonumber \\
& &
+ 
\left(
\begin{array}{cc}       
0 & -1/a_5 \\ 
0 & 0 \\
\end{array}
\right)
\delta_{s,t+1}
+
\left(
\begin{array}{cc}       
0   & 0 \\ 
\mu & 0 \\
\end{array}
\right)
\delta_{sN}
\delta_{t1}
+
\left(
\begin{array}{cc}       
0 & \mu \\ 
0 & 0 \\
\end{array}
\right)
\delta_{tN}
\delta_{s1}
\Bigr] 
\left(
\begin{array}{c}       
\psi_L(n,t) \\ 
\psi_R(n,t) \\
\end{array}
\right). \nonumber\\ 
\end{eqnarray}
Thus $\gamma_4 D_5(\mu)$ takes the form 
\begin{eqnarray}
\left(
\begin{array}{ccccccc}
C^\dagger& B_5      & 0         & \cdots & \cdots & 0      & \mu    \\
 B_5     & -C       & -1/a_5   & \cdots & \cdots & 0      & 0      \\
 0       & -1/a_5  & C^\dagger & B_5    & \cdots & \cdots & 0      \\
 \vdots  &      0   &     B_5   &  -C    & -1/a_5& \cdots & 0      \\
 \vdots  &  \vdots  &  \vdots   & \ddots & \ddots & \cdots & \vdots \\
 0       &    0     &     0     &   0    & -1/a_5& C^\dagger & B_5 \\       
 \mu     &    0     &     0     &   0    &   0    & B_5    &  -C  
\end{array} 
\right),
\end{eqnarray} 
where 
$ C^\dagger = \sum_{\mu} \sigma^\dagger _\mu C_\mu(m,n)  $ and 
$-C = \sum_{\mu} \sigma_\mu C_\mu(m,n)$. Assuming the four dimensional
space consists of $L^4$ sites and the color and flavor groups are 
$SU(N_c)$ and $SU(N_f)$, each block $C$, $B_5$ etc. above is 
$q \times q$ matrix with $q=2 N_C N_f L^4 $. 
The determinant of this matrix is evaluated following the technique
developed in Ref. \cite{neu3,gibbs}. 
Moving the first $q$ columns to the last, $\gamma_4 D_5 $ takes the
from 
\begin{eqnarray} 
& &\left(
\begin{array}{ccccccc}
A_1     & 0    & 0     &\cdots    & 0        & C_1      & B_1   \\
B_2     & A_2  & 0     &\cdots    & 0        & 0        & B_2   \\
C_3     & B_3  & A_3   &0         &\cdots    & \cdots   & 0     \\
 0      & C_4  & B_4   & A_4      & 0        & \cdots   & \vdots\\
 \vdots &\vdots&\vdots &\ddots    &\ddots    & \cdots   & 0     \\
 0      &    0 & 0     &C_{2N_5-1}&B_{2N_5-1}&A_{2N_5-1}& 0     \\  
 0      &\cdots& \cdots&   0      &C_{2N_5}  &B_{2N_5}  & A_{2N_5}  
\end{array} 
\right)
=
\left(
\begin{array}{ccccc}
\alpha_1 &       0  &\cdots    & 0            & \beta_1     \\
\beta_2  & \alpha_2 & 0        &\cdots        & 0           \\
 0       & \beta_3  & \alpha_3 &0             &\cdots       \\
  \vdots &\vdots    &\vdots    &\ddots        & 0       \\
 \cdots  & \cdots   & 0        &\beta_{N_5}  & \alpha_{N_5}  
\end{array} 
\right)
\nonumber  \\
& &\alpha_i = 
\left(
\begin{array}{cc}
A_{2i-1} &       0  \\
B_{2i}   & A_{2i} 
\end{array} 
\right),\,\,\,
\beta_i = 
\left(
\begin{array}{cc}
C_{2i-1} & B_{2i}   \\
  0      & C_{2i} 
\end{array} 
\right). 
\label{eqn:mat1}
\end{eqnarray}
Decomposing the matrix (\ref{eqn:mat1})  
into the product of the two matrices as  
\begin{eqnarray} 
& &\left(
\begin{array}{ccccc}
\alpha_1 &       0  & \cdots    & 0      & \beta_1   \\
\beta_2  & \alpha_2 & 0         & \cdots & 0         \\
 0       & \beta_3  & \alpha_3  & \cdots & \vdots    \\
 \vdots  & \vdots   & \vdots    & \ddots & \vdots    \\
 0       & \cdots   & \cdots    &\beta_{N_5}& \alpha_{N_5}  
\end{array} 
\right)
= 
\left(
\begin{array}{ccccc}
\alpha_1 &       0  &\cdots    & 0            & 0          \\
\beta_2  & \alpha_2 & 0        & \cdots       & 0          \\
 0       & \beta_3  & \alpha_3 & 0            & \vdots     \\
 \vdots  & 0        & 0        & \ddots       & \vdots     \\
 0       & \cdots   &  0       & \beta_{N_5} & \alpha_{N_5}  
\end{array} 
\right)
\nonumber \\
& &\times \left(
\begin{array}{ccccc}
 1      & 0      & \cdots & \cdots  & -v_1       \\
 0      & 1      & 0      & \cdots  & -v_2       \\
 \vdots & 0      & \ddots & \vdots  &  \vdots    \\
 \vdots & \vdots & \vdots & 1       & -v_{N_5-1} \\  
 0      &\cdots  & \cdots & 0       & 1-v_{N_5}  
\end{array} 
\right), \,\,\,\,\,
\begin{array}{c}
v_1=-\alpha^{-1}_1 \beta_1 \\  
-\beta_iv_{i-1}-\alpha_i v_i =0 
\end{array}, 
\label{eqn:mat2}
\end{eqnarray} 
its determinant is given by 
\begin{eqnarray} 
\prod^{N_5}_{i=1} \det \alpha_i \det (1-v_{N_5})  
\end{eqnarray}
and the relations at the end of eq. (\ref{eqn:mat2}) 
leads to 
\begin{eqnarray} 
v_{N_5} = 
( -\alpha^{-1}_{N_5} \beta_{N_5} )\cdots 
( -\alpha^{-1}_{1} \beta_{1} ) .
\end{eqnarray}
In the case of our $\gamma_4 D_5$, 
\begin{eqnarray} 
& &\alpha_1 = \cdots = \alpha_{N_5-1}=  
\left(
\begin{array}{cc}
B_5  &       0  \\
-C   & -1/a_5  
\end{array} 
\right),\,\,\,
\alpha_{N_5}=  
\left(
\begin{array}{cc}
B_5  &       0  \\
-C   & \mu   
\end{array} 
\right),
\nonumber \\
& &\beta_1 =  
\left(
\begin{array}{cc}
-1/a_5  & C^\dagger   \\
  0     & B_5 
\end{array} 
\right),\,\,\,
\beta_2 = \cdots \beta_{N_5}= 
\left(
\begin{array}{cc}
  \mu & C^\dagger    \\
  0   & B_5
\end{array} 
\right)
\end{eqnarray}   
and for the later convenience we introduce the transfer matrix $T$ and the 
Hamiltonian $H_5$ as 
\begin{eqnarray} 
-\alpha^{-1} \beta = \gamma_4 T^{-1} \gamma_4,\,\,\, T=e^{-a_5 H_5},\,\,\, 
T=
\left(
\begin{array}{cc}
  B^{-1}_5/a_5      & B^{-1}_5C    \\
  B^{-1}_5C^\dagger & a_5 C^\dagger B^{-1}_5 C + a_5 B_5
\end{array} 
\right) 
\end{eqnarray}
The final expression is 
\begin{eqnarray} 
\det \gamma_4 D_5(\mu) &=& (-1)^{q(N_5-1)}(a_5)^{-qN_5} (\det B_5)^{N_5} 
\nonumber \\
& &\det 
\Bigl\{
\left(
\begin{array}{cc}
 -a_5 \mu & 0 \\
 0        & 1
\end{array} 
\right) 
-T^{-N_5} 
\left(
\begin{array}{cc}
 1 & 0 \\
 0 & -a_5 \mu
\end{array} 
\right)
\Bigr\},  
\end{eqnarray}
where $(-1)^{q(N_5-1)} $ arises when the first $q$ columns are 
moved into the last column
and eq. (\ref{eqn:pv}) is given by 
\begin{eqnarray} 
\det\frac{1}{2}\{ 1-\gfive \tanh(\frac{1}{2} N_5 a_5 H_5)  \} .
\end{eqnarray}
Taking the limit $N_5 \rightarrow \infty $, 
\begin{eqnarray} 
\tanh(\frac{1}{2} N_5 a_5 H_5) \rightarrow 
\varepsilon(a_5 H_5 )= \frac{a_5 H_5 }{\sqrt{(a_5H_5)^2 }} .
\end{eqnarray}
In the limit $a_5 \rightarrow 0$, 
\begin{eqnarray} 
a_5H_5 \rightarrow H=
\left(
\begin{array}{cc}
 B          & -C \\
 -C^\dagger & -B  
\end{array} 
\right)
=-\gfive X,\,\,\
X=\left(
\begin{array}{cc}
 B          & -C \\
 C^\dagger  & B  
\end{array} 
\right),
\end{eqnarray}
where $X$ is the Wilson-Dirac operator on a lattice. 
Therefore in this limit, eq. (\ref{eqn:pv}) is (up to constant) 
\begin{eqnarray} 
\det D , \quad D=\frac{1}{a}
\Bigl( 
1- \gfive \frac{H}{\sqrt{H^2}} 
\Bigr)
\label{eqn:gwdirac}
\end{eqnarray} 
which allows the path integral expression over the fermion fields 
as eq. (\ref{eqn:z}). 
The Ginsparg-Wilson relation is reduced to $ (H/\sqrt{H^2})^2=1$ for 
Dirac operators of the form eq. (\ref{eqn:gwdirac}), which is 
in fact satisfied by definition \cite{neu3}.     


Now we discuss the weak coupling expansion of the Dirac operator: 
\begin{eqnarray}
  aD &=& 1 + X \frac{1}{\sqrt{X^\dagger X }} , \\
  X_{nm} &=& \gamma_\mu C_\mu(n,m)+B(n,m) - \frac{1}{a} M_0 \delta_{nm} .
\end{eqnarray}
The Wilson-Dirac operator $X$ 
\begin{equation}
  X_{nm} = 
\int \frac{d^4 p}{(2\pi)^4} \frac{d^4 q}{(2\pi)^4} 
e^{i a (q n - p m)} X(q,p) 
\end{equation}
may be expanded as 
\begin{equation}
  X(q,p)=X_0(p) \, (2\pi)^4 \delta^4(q-p) 
        +X_1(q,p) + X_2(q,p) + {\cal O}(g^3) ,
\end{equation}
where
\begin{eqnarray}
X_0(p)&=& 
\frac{i}{a} \gamma_\mu \sin a p_\mu 
+ \frac{r}{a} \sum_\mu \left(1-\cos a p_\mu \right) 
-\frac{1}{a} M_0 ,\\
&& \nonumber\\
X_1(q,p)&=& \int \frac{d^4 k}{(2\pi)^4}  (2\pi)^4 \delta(q-p-k)
\, g A_\mu(k) \, V_{1 \mu}\left(p+\frac{k}{2}\right) , \\
&& \nonumber\\
X_2(q,p)&=& 
\int \frac{d^4 k_1}{(2\pi)^4} \frac{d^4 k_2}{(2\pi)^4} 
(2\pi)^4 \delta(q-p-\sum k_i ) 
\times \nonumber\\
&& \qquad \qquad
\, \frac{g^2}{2} A_\mu(k_1) A_\mu(k_2) \, 
V_{2\mu}\left( p+\frac{\sum k_i}{2} \right) .
\end{eqnarray}
The vertex functions are given explicitly as 
\begin{eqnarray}
V_{1 \mu}\left(p+\frac{k}{2}\right) 
&=& i \gamma_\mu \cos a \left(p_\mu+\frac{k_\mu}{2}\right) 
+ r \sin a \left(p_\mu+\frac{k_\mu}{2}\right)  \\
&=& \frac{\partial}{\partial p_\mu} X_0 \left(p+\frac{k}{2}\right) ,
\nonumber\\
V_{2\mu}\left( p+\frac{\sum k_i}{2} \right) 
&=&
-i \gamma_\mu  a \sin a \left(p_\mu+\frac{k_\mu}{2}\right) 
+ a r \cos a \left(p_\mu+\frac{k_\mu}{2}\right)  . 
\end{eqnarray}

Let us assume that $1/\sqrt{ X^\dagger X} $ can be expanded  in 
the following form, 
\begin{eqnarray} 
\frac{1}{\sqrt{ X^\dagger X}}(p,q) = 
\left(\frac{1}{\sqrt{ X^\dagger X}}\right)_0(p,q) 
+ Y_1(p,q) + Y_2(p,q)+ \cdots .
\end{eqnarray}
Then it should satisfy  
\begin{eqnarray}
\int_q \int_s \, (X^\dagger X)(p,q)
\left(\frac{1}{\sqrt{ X^\dagger X}}\right)(q,s)
\left(\frac{1}{\sqrt{ X^\dagger X}}\right)(s,t) =\delta(p-t) .
\label{eqn:eq1}
\end{eqnarray}
We have made use of the abbreviations as 
$\int \frac{d^4 k}{(2\pi)^4} = \int_k$ and 
$\delta(q-p)= (2\pi)^4 \delta^4(q-p)$.
In the first non-trivial order, the above equation reads
\begin{eqnarray} 
&& \int_s \, \Bigl(\sqrt{X^\dagger X} \Bigr)_0(p,s) Y_1(s,t) 
+  \int_q \int_s \Bigl(X^\dagger X \Bigr)_0 (p,q) Y_1(q,s) 
   \Bigl(\frac{1}{\sqrt{ X^\dagger X}}\Bigr)_0(s,t)
\nonumber \\
& & +\int_{q} \Bigl(X^\dagger X \Bigr)_1 (p,q)
\Bigl(\frac{1}{ X^\dagger X }\Bigr)_0(q,t)=0 .
\label{eqn:first}
\end{eqnarray}
Since $\left(\sqrt{X^{\dagger} X}\right)_0 (p,q)$ is completely
diagonal with respect to the momentum, spinor and color indices, 
\begin{equation}
\left( \sqrt{X^{\dagger} X} \right)_0(p,q)
= \omega(p)\delta(p-q) , \quad 
\end{equation}
where
\begin{equation}
a \, \omega(p)= \sqrt{ 
  \sin^2 a p_\mu 
+ \left( \sum_\mu(1-\cos a p_\mu) - M_0 \right)^2 }  > 0 ,
\end{equation}
Eq. (\ref{eqn:first}) is reduced to the relation between 
the elements $Y_1(p,t)$ and $(X^{\dagger} X)_1(p,t)$,
leading to the solution \cite{takahashi}, 
\begin{eqnarray}
Y_1(p,t)= 
-\frac{1}{\omega(p)\omega(t)}
\Bigl\{ 
\frac{1}{\omega(p)+\omega(t)} 
\Bigr\}
(X^\dagger X)_1(p,t).
\end{eqnarray}
Similar procedure yields the expansion of the Dirac operator 
\begin{eqnarray}
\label{eq:weak-coupling-expansion-p}
D(p,q)
&=& D_0(p) \, \delpq + V(p,q) \\
V(p,q) &=& 
\Bigl\{ 
\frac{1}{\omega(p)+\omega(q)} 
\Bigr\}
\Bigl[
X_1(p,q)
-
\frac{X_0(p)}{\omega(p)} X^\dagger_1(p,q) 
\frac{X_0(q)}{\omega(q)}
\Bigr] 
\nonumber \\
&+ &\Bigl\{ 
\frac{1}{\omega(p)+\omega(q)} 
\Bigr\}
\Bigl[
X_2(p,q)
-
\frac{X_0(p)}{\omega(p)} X^\dagger_2(p,q) 
\frac{X_0(q)}{\omega(q)}
\Bigr] 
\nonumber \\ 
&+ &
\int_{t} 
\Bigl\{\frac{1}{\omega(p)+\omega(q)}\Bigr\} 
\Bigl\{\frac{1}{\omega(p)+\omega(t)}\Bigr\}
\Bigl\{\frac{1}{\omega(t)+\omega(q)}\Bigr\} 
\times
\nonumber \\ 
&& \quad
\Bigl[ \
-X_0(p) X^\dagger_1(p,t)X_1(t,q)  
\nonumber \\
&& \qquad
-X_1(p,t)X^\dagger_0(t) X_1(t,q)    
-X_1(p,t)X^\dagger_1(t,q)X_0(q)  
\nonumber \\
&& \qquad
+\frac{\omega(p) +\omega(t) +\omega(q)}{\omega(p) \omega(t) \omega(q)}
X_0(p) X^\dagger_1(p,t) X_0(t) X^\dagger_1(t,q) X_0(q)  \
\Bigr] + \cdots . 
\label{eqn:dex}
\end{eqnarray}

This expansion may be obtained through the integral 
representation of the inverse square root of $X^\dagger X$:
\begin{equation}
\frac{1}{\sqrt{X^\dagger X}} 
= 
\int^\infty_{-\infty} \frac{dt}{\pi} \frac{1}{t^2 + X^\dagger X} .
\end{equation}
Up to the second order, we have
\begin{eqnarray}
\frac{1}{\sqrt{X^\dagger X}} 
&=&
\int^\infty_{-\infty} \frac{dt}{\pi} \frac{1}{t^2 + X_0^\dagger X_0}
- \int^\infty_{-\infty} \frac{dt}{\pi} 
\frac{1}{t^2 + X_0^\dagger X_0}
\left(
X_0^\dagger X_1 + X_1^\dagger X_0 
\right)
\frac{1}{t^2 + X_0^\dagger X_0} \nonumber\\
&& 
- \int^\infty_{-\infty} \frac{dt}{\pi} 
\frac{1}{t^2 + X_0^\dagger X_0}
\left(
X_0^\dagger X_2 + X_2^\dagger X_0 + X_1^\dagger X_1 
\right)
\frac{1}{t^2 + X_0^\dagger X_0} \nonumber\\
&&
+ \int^\infty_{-\infty} \frac{dt}{\pi} 
\frac{1}{t^2 + X_0^\dagger X_0}
\left(
X_0^\dagger X_1 + X_1^\dagger X_0 
\right)
\frac{1}{t^2 + X_0^\dagger X_0} 
\left(
X_0^\dagger X_1 + X_1^\dagger X_0 
\right)
\frac{1}{t^2 + X_0^\dagger X_0} .
\nonumber\\
\end{eqnarray}
Noting that $(t^2+X_0^\dagger X_0)^{-1}$ is diagonal in spinor space
and it commutes with $X_0$, we obtain
\begin{eqnarray}
a D &=& a D_0  
\nonumber\\
&& 
+ \int^\infty_{-\infty} \frac{dt}{\pi} 
\frac{1}{t^2 + X_0^\dagger X_0}
\left( t^2 X_1 - X_0  X_1^\dagger X_0 
\right)
\frac{1}{t^2 + X_0^\dagger X_0} \nonumber\\
&& 
+ \int^\infty_{-\infty} \frac{dt}{\pi} 
\frac{1}{t^2 + X_0^\dagger X_0}
\left( t^2 X_2 - X_0  X_2^\dagger X_0 
\right)
\frac{1}{t^2 + X_0^\dagger X_0}  \nonumber\\
&&
- \int^\infty_{-\infty} \frac{dt}{\pi} \, t^2 \, 
\frac{1}{t^2 + X_0^\dagger X_0}
\left( X_1
\right)
\frac{1}{t^2 + X_0^\dagger X_0} 
\left(
X_0^\dagger X_1 + X_1^\dagger X_0 
\right)
\frac{1}{t^2 + X_0^\dagger X_0} \nonumber\\
&&
-\int^\infty_{-\infty} \frac{dt}{\pi} \, t^2 \, 
\frac{1}{t^2 + X_0^\dagger X_0}
\left(
X_0 X_1^\dagger 
\right)
\frac{1}{t^2 + X_0^\dagger X_0} 
\left(
X_1 
\right)
\frac{1}{t^2 + X_0^\dagger X_0} \nonumber\\
&&
+ \int^\infty_{-\infty} \frac{dt}{\pi} 
\, \left(X_0\right) \, 
\frac{1}{t^2 + X_0^\dagger X_0}
\left(
X_1^\dagger X_0 
\right)
\frac{1}{t^2 + X_0^\dagger X_0} 
\left(
X_1^\dagger X_0 
\right)
\frac{1}{t^2 + X_0^\dagger X_0} + \cdots 
\end{eqnarray}
Going to the momentum space, the integration over the parameter $t$
can be performed explicitly 
and we obtain Eq.~(\ref{eq:weak-coupling-expansion-p}).

The tree level propagator is given by 
\begin{eqnarray}
D^{-1}_0(p)=
\frac{- i \sum_{\mu} \gamma_\mu \sin a p_\mu}
     {2 \left(\omega(p) + b(p) \right)}
+ \frac{a}{2}, 
\end{eqnarray}
\begin{equation}
b(p)= 
\frac{r}{a} \sum_\mu \left(1-\cos a p_\mu \right) -\frac{1}{a} M_0 .
\end{equation}
Since $\omega(p)$ is positive definite, the pole of the propagator 
occurs when $\omega(p)+b(p)=0$, 
which is fulfilled only if $\sin^2 a p_\mu=0 $ and $b(p)<0 $. 
Therefore for $-2r < M_0 <0 $ the propagator exhibits a massless 
pole only when $p_\mu = 0$. 


Now using the expansion Eq. (\ref{eqn:dex}) we compute  
$-a \tr  \varepsilon \gfive D$. (Here the local transformation 
is treated so that $\varepsilon$ is space-dependent.) 
The non-vanishing contribution comes only from the term containing 
$X_0(p) X^\dagger_1(p,t) X_0(t) X^\dagger_1(t,q) X_0(q)$ . 
We obtain 
\begin{eqnarray}
 -a \tr \varepsilon \gfive D &=& a^4  \int_{pq} \varepsilon_m 
e^{i(p-q)am} g^2
\tr \{A_\mu(p) A_\nu(-q)\}  f_{\mu\nu}(p,q), 
\label{eqn:anomaly}
\\
\nonumber\\
f_{\mu\nu}(p,q) &=& \int_{k} \Omega(p+k,k,k+q) \times 
\nonumber\\
&& \qquad 
\tr \gfive 
X_0(k+p) \partial_\mu X_0^\dagger(k+p/2) X_0(k) 
       \partial_\nu X_0^\dagger(k+q/2) X_0(k+q) ,
\nonumber \\
\\
\Omega(p,k,q)&=& 
\Bigl\{ \frac{1}{\omega(p)+\omega(q)} \Bigr\}
\Bigl\{ \frac{1}{\omega(p)+\omega(k)} \Bigr\}
\Bigl\{ \frac{1}{\omega(k)+\omega(q)} \Bigr\}
\Bigl\{ 
\frac{\omega(p)+\omega(k)+\omega(q)}{\omega(p)\omega(k)\omega(q)}
\Bigr\} .
\nonumber \\
\end{eqnarray}
Then using the propagator $S(p,p_5)$ defined in five dimensional 
momentum space $(p_\mu, p_5) \in  T^4\times R$, 
\begin{eqnarray}
S(p,p_5) ^{-1}= X_0(p) + i \gamma_5 p_5 
\end{eqnarray}
$f_{\mu\nu}(p,q)$ is rewritten as 
\begin{eqnarray}
f_{\mu\nu}(p,q)= 2i \int_{(k_i,k_5) \in T^4\times R } 
\tr 
\{ 
S(k+p, k_5) \partial_{\mu} S^{-1}(k+p/2,k_5) 
\nonumber \\
S(k,k_5) \partial_{\nu} S^{-1}(k+q/2,k_5) 
S(k+q,k_5) \partial_{5} S^{-1}(k,k_5) 
\} .
\label{eqn:f}
\end{eqnarray}
Now it is easy to verify the structure 
$f_{\mu\nu}(p,q)=
p_\rho q_\tau 
\partial_\rho \partial_\tau f_{\mu\nu}(p,q)|_{p,q=0} + {\cal O}(a)$, 
and $\partial_\rho \partial_\tau f_{\mu\nu}(p,q)|_{p,q=0} \propto 
\epsilon_{\rho\mu\tau\nu} $, leading to the 
anomaly $g^2 c(M_0,r) \epsilon_{\rho\mu\tau\nu}F_{\rho\mu}F_{\tau \nu}$ 
when inserted in eq. (\ref{eqn:anomaly}) with the coefficient 
\begin{eqnarray}
c(M_0,r) &=& 2i \frac{1}{2^2 5!}
\epsilon_{\mu_1\mu_2\mu_3\mu_4\mu_5}  
\int_{T^4 \times R }
tr 
\Bigl\{ 
L_{\mu_1}L_{\mu_2}L_{\mu_3}L_{\mu_4}L_{\mu_5}
\Bigr\},
\label{eqn:c} \\
& & L_{\mu}=S \partial_\mu S^{-1}.
\nonumber 
\end{eqnarray}

This expression shows that $c(M_0,r) $ is invariant  
by a momentum-dependent, continuous change of the scale 
of the propagator $S$: $S(p) \rightarrow \Omega(p)S(p)$ 
as long as $ \Omega(p) \neq 0$ in $T^4 \times R $. 
It allows to replace 
$S$ and $S^{-1}$ in eq. (\ref{eqn:c}) 
with 
$V$ and $V^{-1}$ where  
\begin{equation}
  V(p,p_5) = N(p,p_5) S(p,p_5), \qquad
N(p,p_5)= \sqrt{ p_5^2 + \omega(p)^2 } .
\end{equation}
$V$ is a mapping from $T^4 \times R$ to $ S^5 $
and $n_5 $ monotonously increases from $-1$ to $1$ as  
$p_5 $ increases from $-\infty$ to $\infty$. This 
allows one to interpret that   
$c(M_0,r) $ is the winding number of the mapping, $V'$, from 
$T^4 $ to $ S^4 \subset S^5 $ derived from $V$ by fixing $p_5$ 
and it is an integer.  
For $M_0 > 0$, $n_0 (\propto b)$ is positive definite 
on $T^4 \times R$ and the image of $T^4$ by $V'$ 
does not cover $S^4 \subset S^5$, leading to the value
$c(M_0,r)=0$. Further studies along this line leads to the value 
$c(M_0,r)=1/16 \pi^2 $ for $-2r <M_0 <0 $ \cite{top}.  
(Note that our convension for the gamma matrices is as 
$ \gamma_1 \gamma_2 \gamma_3 \gamma_4 \gamma_5 = - 1 $. )

This value is also obtained by evaluating the difference 
$ \Delta c = c(\epsilon,r) - c(-\epsilon ,r)$ in the limit 
$\epsilon \rightarrow 0 $ \cite{so}.  
A continuous and momentum dependent deformation of $M_0$ does not 
change $c(M_0,r)$ as long as $N \neq 0 $ in $T^4 \times R $, stemming 
from the fact of $c(M_0,r)$ being the winding number. Here
the surfaces $M_0 = \pm \epsilon  $ are deformed to 
$m_\pm(p) $ defined on $T^4$, where 
$m_\pm(p)=0$ for $ \sum_{\mu=1}^{4} p^2_\mu \gsimeq \delta^2  $ 
and   
$m_\pm(p)=\pm \epsilon$ for 
$\sum_{\mu=1}^{4} p^2_\mu \lsimeq \delta^2$. Then the difference 
$\Delta c$ arises from the integration in the vicinity of the 
center of the Brillouin zone in eq. (\ref{eqn:c}). For sufficiently 
small $\delta$ (still larger than $\epsilon$ ), $S$ and $S^{-1}$ may be 
replaced by their continuum expressions, leading to 
\begin{eqnarray} 
\Delta c =  2 \int_{|p| < \delta} 
\frac{2\epsilon}{(N^3)} \rightarrow  -\frac{1}{16\pi^2}
\end{eqnarray}
by taking the limit $\epsilon\rightarrow0 $ for a finite small $\delta$. 

Thus we obtain the final result of the axial $U(1)$ anomaly as 
\begin{eqnarray} 
-a \tr \left\{ \varepsilon \gfive D \right\} 
=  \frac{g^2}{32\pi^2} N_f \int d^4 x \, \varepsilon(x) 
   \epsilon_{\rho\mu\tau\nu} F_{\rho\mu}^a(x) F_{\tau \nu}^a(x) .
\end{eqnarray}
The same analysis can be done in two dimensions where the anomaly comes 
from the second term in eq. (\ref{eqn:dex}). The result is
\begin{eqnarray} 
-a \tr \left\{ \varepsilon \gfive D \right\} 
= \frac{g}{2\pi} N_f \int d^2 x \, \varepsilon(x) \epsilon_{\mu\nu}
F_{\mu\nu}(x) .
\end{eqnarray}

We have discussed the weak coupling expansion of 
the lattice QCD with a Dirac operator satisfying the Ginsparg-Wilson 
relation using the expression of Ref. \cite{neu1}. We confirmed that the
anomalous behavior of the fermion partition function under the axial
$U(1)$ transformation of Ref. \cite{lu} is expressed in
the form of the topological charge density for slowly varying 
perturbative gauge fields, which supplements the earlier calculation 
in Ref. \cite{gw}

Narayanan, Vranas and Singleton Jr.
have studied numerically the relation between the index of the Dirac operator 
${\rm Tr} \left( H/\sqrt{H^2} \right)$ and the topological charge 
of lattice gauge field \cite{index-topological-charge}.
An interesting possibility for studying this type of Dirac operator
in numerical simulation was considered recently by Chiu \cite{chiu}.


YK would like to thank H.~Neuberger, T.~Kugo and M.~Fukuma for discussions.
YK is also grateful to T.-W.~Chiu and S.~Zenkin for discussions.
YK is supported in part by Grant-in-Aid for Scientific Research from
Ministry of Education, Science and Culture(\#10740116,\#10140214).
AY would like to thank Ken-ichi Izawa, Teruhiko Kawano, 
Masatoshi Sato and Tomohiko Takahashi for useful discussions. 



\begin{thebibliography}{99}
\bibitem{ex1}
This was shown in Ref. \cite{shamir} by studying the free fermion 
propagator for the action (\ref{eqn:shamia}). 
The eigenvalues of the mass matrix were explicitly obtained 
in Ref. \cite{neu3}.
\bibitem{takahashi}
The following relation may be also useful;
\begin{eqnarray}
\frac{1}{\sqrt{A+B}} =\frac{1}{\Gamma(1/2)} 
\int^{\infty}_{0} dt t^{-1/2} e^{-tA}e^{tA}e^{-t(A+B)}. 
\end{eqnarray} 
One of us (AY) would like to thank Tomohiko Takahashi 
for the discussion on this point.
\bibitem{top}
For example, when we sweep the $T^4$ with the hypersurface of constant $b$ 
starting from the center of the Brillouin zone,   
the region within the first coner of the Brillouin zone covers 
$S^4 \subset S^5$ completely by the mapping $V'$ 
and the image of the remaining region covers and
uncovers certain region in the upper half side of $S^4$ 
centered by the north pole in a oscillatory way 
as its inverse image goes away from the center of the Brillouin zone, 
and finially converges to the north pole.
\end{thebibliography}
\end{document}